\documentclass[a4paper,10pt,conference]{IEEEtran}

\usepackage[utf8]{inputenc}
\usepackage[T1]{fontenc}
\usepackage{textcomp}
\usepackage{fixltx2e}
\usepackage{microtype}
\usepackage{calc}
\usepackage[normalem]{ulem}
\usepackage{authblk}

\usepackage{lipsum}
\usepackage{soul}

\usepackage[binary-units=true]{siunitx}
\usepackage{silence}
\AtBeginDocument{
    \DeclareSIUnit\decibel{dB}
}

\usepackage{csquotes}
\usepackage[backend=biber,style=ieee,doi=false,isbn=false,mincitenames=1,maxcitenames=2]{biblatex}
\DeclareFieldFormat{sentencecase}{#1} 
\DeclareFieldFormat{titlecase}{#1} 
\usepackage{xpatch}
\xpatchbibmacro{textcite}{\addspace}{\addnbspace}{}{}
\xpatchbibmacro{Textcite}{\addspace}{\addnbspace}{}{}
\DefineBibliographyStrings{english}{
    andothers = et~al\adddot\addspace
}

\usepackage{amsmath}

\usepackage{amssymb}
\usepackage{amsfonts}
\usepackage{amstext}
\usepackage{amsthm}

\usepackage{booktabs}
\usepackage{multirow}
\usepackage{url}

\usepackage[inline]{enumitem}

\usepackage[caption=false,font=footnotesize]{subfig}
\usepackage[pdftex]{graphicx}
\DeclareGraphicsExtensions{.pdf,.png,.jpg,.tikz}
\usepackage{tikz}
\usetikzlibrary{arrows}
\usetikzlibrary{calc}
\usetikzlibrary{chains}
\usetikzlibrary{scopes}

\usepackage[american]{babel}
\usepackage{hyphenat}

\usepackage[]{algorithmic}
\usepackage[ruled,vlined]{algorithm2e}

\usepackage[capitalize,noabbrev]{cleveref}
\crefformat{footnote}{#2\footnotemark[#1]#3}
\Crefformat{figure}{#2Fig.~#1#3}
\Crefformat{equation}{#2Eq.~(#1)#3}

\usepackage{todonotes}

\usepackage[color]{changebar}

\usepackage{floatpag} 

\usepackage{xcolor}

\usepackage{listings}
\usepackage{color}
\usepackage[hang]{footmisc}
\usepackage{afterpage} 

\usepackage{optidef}
\usepackage{bm}
\usepackage{tabularx}

\RequirePackage{xstring}
\RequirePackage{xparse}
\RequirePackage[]{acro}
\NewDocumentCommand\acrodef{mO{#1}mG{}}{\DeclareAcronym{#1}{short={#2}, long={#3}, #4}}
\acrodef{5G}{Fifth Generation}
\acrodef{ANN}{Artificial Neural Network}
\acrodef{AUC}{Area under the curve}
\acrodef{BCE}{Binary Crossentropy}
\acrodef{CNN}{Convolutional Neural Network}
\acrodef{DFT}{Discrete Fourier Transform}
\acrodef{FPR}{False Positive Rate}
\acrodef{GDPR}{General Data Protection Regulation}
\acrodef{GRU}{Gated Recurrent Unit}
\acrodef{HAR}{Human Activity Recognition}
\acrodef{IQR}{Interquartile Range}
\acrodef{LSTM}{Long Short-Term Memory}
\acrodef{OSM}{Open Street Map}
\acrodef{OS}{Operating System}
\acrodef{RELU}{Rectified Linear Unit}
\acrodef{RNN}{Recurrent Neural Network}
\acrodef{ROC}{Receiver operating characteristic}
\acrodef{SAS}{Sensitivity at Specificity}
\acrodef{SF}{Sensor-based Fusion}
\acrodef{SGD}{Stochastic Gradient Descent}
\acrodef{TPR}{True Positive Rate}

\begin{document}

\title{A Realistic Cyclist Model for SUMO Based on the SimRa Dataset}
\author[1]{Ahmet-Serdar Karakaya}
\author[1]{Konstantin Köhler}
\author[2]{Julian Heinovski}
\author[2]{Falko Dressler}
\author[1]{David Bermbach}
\affil[1]{\textit{TU Berlin \& ECDF, Mobile Cloud Computing Research Group}\\
\{ask,koko,db\}@mcc.tu-berlin.de}
\affil[2]{\textit{TU Berlin, Telecommunication Networks}\\
\
\{heinovski,dressler\}@ccs-labs.org}
\date{}
\maketitle

\begin{abstract}
Increasing the modal share of bicycle traffic to reduce carbon emissions, reduce urban car traffic, and to improve the health of citizens, requires a shift away from car-centric city planning.
For this, traffic planners often rely on simulation tools such as SUMO which allow them to study the effects of construction changes before implementing them.
Similarly, studies of vulnerable road users, here cyclists, also use such models to assess the performance of communication-based road traffic safety systems.
The cyclist model in SUMO, however, is very imprecise as SUMO cyclists behave either like slow cars or fast pedestrians, thus, casting doubt on simulation results for bicycle traffic.
In this paper, we analyze acceleration, velocity, and intersection left-turn behavior of cyclists in a large dataset of real world cycle tracks.
We use the results to derive an improved cyclist model and implement it in SUMO.
\end{abstract}


\section{Introduction}
\label{sec:intro}
Active transportation modes such as cycling provide health benefits, alleviate traffic congestion, and reduce air pollution~\cite{goetschi2016cycling}.
In practice, however, cyclists often face a car-centric traffic infrastructure which has a significant impact on their (perceived) safety and also affects the attractiveness of cycling routes~\cite{karakaya2020simra, pedroso2016bicycle,aldred2018predictors}.
Changing this infrastructure to better accommodate cyclists and pedestrians requires significant planning efforts of city planners and traffic engineers.
Similarly, road traffic safety systems for vulnerable road users are often assessed using simulation.
Particularly the interaction with cars is relevant when it comes to V2X-based safety systems for cyclists~\cite{oczko2020integrating}.
Many of these studies rely on the open source simulation platform SUMO\footnote{https://www.eclipse.org/sumo/} (Simulation of Urban Mobility), which allows them to study the effects of infrastructure changes before implementing them on the streets.

In SUMO, vehicles and their dynamics are simulated individually~\cite{lopez2018microscopic}.
Unfortunately, the cyclist model is not particularly realistic -- cyclists can either be modeled to behave as slow cars or as fast pedestrians.
Several studies have already improved the bicycle model of SUMO.
For instance, \textcite{kaths2016integration} investigated the intersection behavior of cyclists using camera traces and transferred findings into SUMO.
Also, \textcite{grigoropoulos2019modelling} improved modeling of bicycle infrastructure at intersections while \textcite{heinovski2019modeling} created a virtual cycling environment to import real bicycle behavior directly into SUMO.
Nevertheless, the current cyclist behavior in SUMO is still rather unrealistic; so far, researchers have devoted much more effort to car models, e.g., \cite{chandler1958traffic,gazis1961nonlinear,gipps1981behavioural, leutzbach1986development,bando1995dynamical,krauss1998microscopic,treiber2000congested,salles2020extending}.
One reason for this is that, until recently, not enough data on real-world cyclist behavior have been available.
Today, crowdsourced data collection approaches such as SimRa\footnote{https://github.com/simra-project/}~\cite{karakaya2020simra} have made thousands of cycle tracks available as open data.

In this paper, we analyze the SimRa dataset regarding acceleration and velocity of cyclists as well as their left-turn behavior in four-way intersections.
We then use our findings to improve the cyclist model in SUMO.
In this regard, we make the following contributions:
\begin{itemize}
    \item We show that SUMO's default bicycle simulation is not realistic (\cref{sec:analysis}),
    \item we improve bicycle simulation in SUMO by deriving new parameters for that vehicle type in SUMO (\cref{sec:concept}),
    \item we develop an intersection model which captures cyclists' left-turn behavior at intersections in a more realistic way (\cref{sec:concept}), and
    \item we compare our improvements to SUMO's default bicycle simulation, using the SimRa dataset as a ground truth (\cref{sec:eval}).
\end{itemize}

\section{Background}
\label{sec:background}
In this section, we give an overview of SUMO (see~\cref{sub:sumo}) and SimRa (see~\cref{sub:simra}), which provided the dataset we used in our work.

\subsection{SUMO}
\label{sub:sumo}

SUMO is an open source traffic simulation tool that offers macroscopic as well as microscopic simulation of vehicle mobility~\cite{lopez2018microscopic}.
SUMO includes models for different types of ``vehicles'', including, among others, cars, bicycles, and even pedestrians.
Due to its large feature set, it has become the de-facto standard for traffic simulation and is used even beyond the transport community, e.g., \cite{beilharz2021towards}.

Traffic scenarios are, among other things, defined by road networks and vehicle traffic.
The road network includes roads and their (sub-)lanes as well as exclusive lanes for cyclists and pedestrians, or road-side infrastructure such as traffic lights.
Furthermore, connections between these lanes and traffic lights can be configured.

When modeling vehicle traffic, users specify demand for a specific road segment per vehicle type and can adjust vehicle-specific parameters of SUMO's simulation model to control their respective behavior.
In general, vehicle parameters are usually specified in the vehicle type declaration (\textit{vType}), applying the changes to all instances of the respective \textit{vType}, e.g., to all cars.
An alternative, however, is to obtain multiple \textit{vType} realizations which typically differ in at least one parameter by using so-called \textit{vTypeDistributions}.
This way, when spawning a new vehicle, SUMO randomly picks a specific \textit{vType} from the \textit{vTypeDistribution} and instantiates the vehicle's parameters accordingly, e.g., cars can thus have individual maximum velocities.

In SUMO, vehicle behavior is, among other things, defined by
\textit{Car Following (CF) models} for the longitudinal kinematic behavior,
\textit{Lane Change (LC) models} for the lateral kinematic behaviour,
and \textit{junction models} for the behavior at junctions and intersections.

Despite including several of these models for cars and trucks, SUMO does not provide a dedicated movement model for cyclists.
Instead, cyclists are simulated by modeling them either as slow cars or fast pedestrians.
Both of these approaches use movement models of the corresponding vehicle type and adapt their respective shape and kinematic characteristics (e.g., velocity and acceleration profiles) to match cyclists.
While this is obviously a rough approximation, it is unlikely to reflect the behavior of real-world cyclists~\cite{grigoropoulos2019modelling}.

\subsection{SimRa}
\label{sub:simra}

SimRa is an open source project started in 2019 which aims to identify hotspots of near miss incidents in bicycle traffic~\cite{karakaya2020simra, temmen2022crowdsourcing}.
For this, the project follows a crowdsourcing approach in which cyclists record their daily rides using a smartphone application available for both Android and iOS.
Today, the project has managed to record more than 65,000 rides, most of them in Germany, approximately half of them in Berlin.

During the ride, SimRa records the GPS trace at 1/3Hz and the motion sensors, i.e., (linear) accelerometer, gyroscope, and rotation vector at 50Hz;
the motion sensor readings are aggregated by calculating a moving average with a window size of 30 and then keeping only every fifth value.
This was done for saving memory, battery, and mobile data usage while still being able to reconstruct the ride and detect near miss incidents.
After the ride, SimRa shows the recorded ride as a route on the map which is then annotated, cropped (for privacy reasons), and uploaded by the user.
In this paper, we only use measured data from the ride files and disregard user-annotated data on near miss incidents.

\section{Related Work}
\label{sec:rw}
In this section, we give an overview of related work on improving intersection behavior (\cref{sec:rw_intersection}) and longitudinal (\cref{sec:rw_longitudinal}) behavior of cyclists in SUMO's simulation models.

\subsection{Intersection Behavior of Cyclists}
\label{sec:rw_intersection}

\textcite{kaths2016integration} aim to address the shortcomings of SUMO's intersection model for cyclists.
For this, they record video footage of an example intersection in Munich and derive cyclist trajectories.
From the set of trajectories, they select one representative trajectory for each combination of start and end points in the intersection and make it available to SUMO via an external API.
While this is a significant improvement in realism over SUMO's intersection model, it is hard to generalize to other intersections and cannot cover the plurality of trajectories chosen by real-world cyclists.


Similar to \textcite{kaths2016integration}, \textcite{grigoropoulos2022traffic} analyze video footage of intersections with the goal of better understanding the intersection behavior of cyclists.
Their focus, however, is not on deriving an improved intersection model but rather on identifying best practices for traffic planners working on intersections with high volumes or cycling traffic.
\textcite{grigoropoulos2019modelling} propose to adjust the default traffic infrastructure inside SUMO to achieve more realistic cyclist behavior at intersections.
Here, they focus on the number and shape of bicycle lanes which, however, are highly specific and differ from intersection to intersection.

\subsection{Longitudinal Behavior of Cyclists}
\label{sec:rw_longitudinal}

\textcite{twaddle2016modeling} examine four models for the longitudinal kinematic behavior of cyclists, i.e., acceleration and velocity.
The first, called Constant Model, is the most simple one and is the SUMO default:
Cyclists accelerate and decelerate at a constant rate until the desired velocity is reached.
This model works well when breaking to a full stop but leads to frequent acceleration jumps between a fixed positive or negative value and zero, which is not realistic cyclist behavior.
In the Linear Decreasing Model, maximum acceleration is reached, when starting the acceleration maneuver and then linearly declines until the desired velocity is reached.
This model is outperformed by all other models.
In the third and fourth models, Polynomial and Two Term Sinusoidal Model, acceleration or deceleration start at zero and then gradually grow over time.
In their paper, \textcite{twaddle2016modeling} analyze the video recordings of 1030 rides in four intersections in Munich, Germany and conclude that the Polynomial Model has overall the most realistic cyclist behavior but is, however, not trivial to implement in SUMO.

A different approach of achieving realistic cycling behavior in SUMO is taken by \textcite{heinovski2019modeling}.
The authors simulate multiple traffic scenarios in which accidents between cars and cyclists occur to investigate the effects of wireless communication between cyclists and other road users in the context of accident prevention.
In order to obtain realistic cycling behavior for SUMO, they set up a novel Virtual Cycling Environment (VCE) featuring an actual bicycle that is connected to the simulation via multiple sensors.
The VCE supports interactive empirical studies in a physically safe environment and allows the authors to record the cyclists' behavior in the form of trajectories.
They use a set of recorded trajectories from different cyclists for emulating realistic cycling behavior inside SUMO to simulate accidents.
Although their approach produces trajectories from cyclists created with an actual bicycle, it does only achieve limited realism, since no other road users were present when recording the trajectories.
Furthermore, deriving a realistic set of trajectories requires a large number of test persons.

\section{Cycling Behavior in SimRa and SUMO}
\label{sec:analysis}
In this section, we analyze real-world cyclists' behavior extracted from the SimRa dataset and compare it to the behavior of SUMO's default bicycle model.
We analyze acceleration and velocity behavior in \cref{sec:analysis_acceleration,sec:analysis_velocity} before discussing left-turn behavior at four-way intersections in \cref{sec:analysis_pathfinding}.
We omit a detailed discussion of the right-turn behavior at intersections, since SUMO's default model does not deviate much from the behavior observed from SimRa's dataset.
When referring to SUMO's bicycle model, we refer to the ``slow car'' model of SUMO as the ``fast pedestrian'' model occasionally leads to poor results and was therefore not considered further.

\begin{table}[b]
\centering
\caption{Most important attributes of entities the SiMRa dataset}
\label{tab:dataset}
\begin{tabular}{ccc}
\toprule
& Total & Used \\
\midrule
\midrule
Rides & \num{57662} & \num{43961} \\
Acceleration Maneuvers & \num{1922087} & \num{140736 } \\
\bottomrule&
\end{tabular}
\end{table}

Aside from the public SimRa datasets~\cite{dataset_simra_set1,dataset_simra_set2} and more recent rides available on GitHub,\footnote{https://github.com/simra-project/dataset} we also used non-public rides which have, for privacy reasons, not been published yet.
\Cref{tab:dataset} summarizes the most important attributes of the dataset that we used.

Since SimRa's dataset stems from crowdsourced smartphone data generation, it suffers from poor sensor quality~\cite{chowdhury2014estimating, usami2018bicycle} as well as heterogeneous hardware and users~\cite{basiri2018impact}.
To achieve the best possible data quality, we tested various pre-processing techniques and filters.
We also conducted an experiment in which sample trajectories were recorded in parallel on several SimRa client devices and compared to a ground truth trajectory recorded by a stand-alone GPS receiver.
In the end, we used a Gaussian Kernel filter for improving location data and a Low Pass filter for the velocity data.
After filtering semantically and syntactically defective files, we used data from \num{43961} rides as input for our analysis scripts.\footnote{https://github.com/simra-project/SimRaXSUMO}

Since most of the data were recorded in Berlin, Germany, all examples in the following focus exclusively on Berlin scenarios.

\subsection{Acceleration}
\label{sec:analysis_acceleration}

\begin{figure}
  \centering
    \includegraphics[width=\columnwidth]{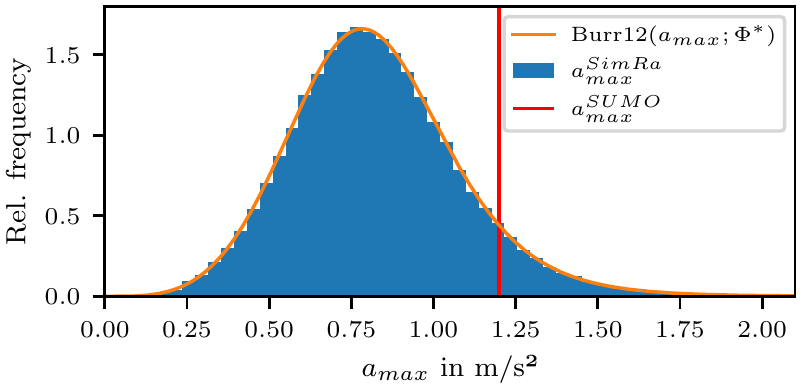}
    \caption{Histogram of the empirical maximum acceleration capabilities of cyclists found in the SimRa dataset and the fitted EMG. The red scalar represents the default value in SUMO.}
    \label{fig:analysis_max_acceleration_dist_fit}
\end{figure}

For analyzing cyclist acceleration, we extracted acceleration maneuvers from the dataset.
For this, we slightly adapted the approach of \cite{ma2016modeling} and found \num{140736} acceleration maneuvers in the cleaned dataset.
Distribution fitting processes showed that the Burr (Type XII) distribution~\cite{burr1942cumulative} Burr($a_{max}; \Phi\mbox{*}$) fits the data best (see also \cref{fig:analysis_max_acceleration_dist_fit}).

Comparing the acceleration capability of actual cyclists (the SimRa dataset) with the default SUMO bicycle model, differences become apparent. By default, SUMO specifies $a_{max}^{SUMO}$ with 1.2m/s\textsuperscript{2}.
This deviates significantly from the findings in the SimRa dataset where only 7.7\% of the acceleration maneuvers are executed with a maximum acceleration of 1.2m/s\textsuperscript{2} or higher.
Furthermore, the empirical distribution is rather wide, indicating a broad variance across different cyclists and cycling situations, which is in stark contrast to SUMO's strategy of choosing a fixed maximum value.

\subsection{Velocity}
\label{sec:analysis_velocity}

\begin{figure}
  \centering
    \includegraphics[width=\columnwidth]
    {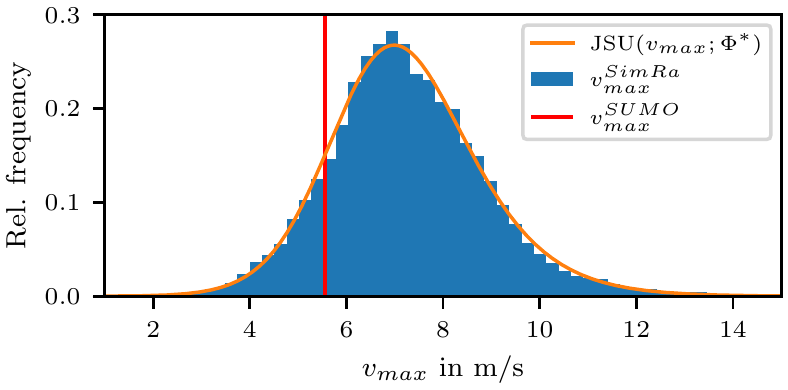}
    \caption{Histogram of the empirical maximum velocity data and the fitted JSU. The red scalar represents the default value in SUMO.}
    \label{fig:analysis_max_velo_dist_fit}
\end{figure}

To gain insights into cyclists' behavior regarding their velocities, we calculate the maximum velocity for each ride file in the cleaned SimRa dataset.
Using distribution fitting, we found that the S\textsubscript{$U$} distribution \textcite{johnson1949systems} JSU($v_{max};\Phi\mbox{*}$) fits the empirical data best (see also \cref{fig:analysis_max_velo_dist_fit}) and is therefore a valid fit for the specification of the empirical distribution of $v_{max}^{SimRa}$.

On the other hand, SUMO sets $v_{max}^{SUMO}$ at 5.56m/s by default.
This deviates significantly from the findings in the SimRa dataset where 86.7\% of the rides have a higher maximum velocity.
Bringing this together with the acceleration findings, real-world cyclists often (but not always) cycle much faster than SUMO cyclists and vary much more in their acceleration behavior.

\subsection{Left-turn Behavior at Intersections}
\label{sec:analysis_pathfinding}

\begin{figure}
\includegraphics[width=\columnwidth]{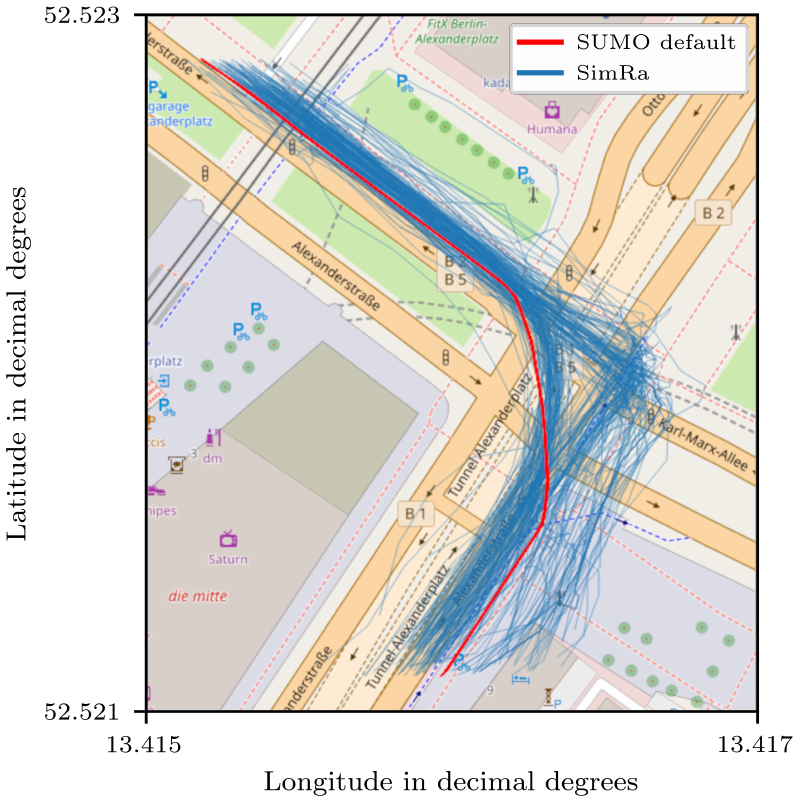}
\caption{
    Qualitative comparison between the SUMO default intersection model and real world data given by SimRa for the intersection between Alexanderstraße and Karl-Marx-Allee in Berlin.
    SimRa shows two distinct left-turn paths (i.e., a direct and an indirect one) whereas SUMO default only models the direct path.
}
\label{fig:analysis_im_traj_default}
\end{figure}

According to the SimRa dataset, cyclists either behave like cars (using the normal road) or pedestrians (using the pedestrian crossing) to take left-turns at intersections.
We call the former a \textit{direct} left turn and the latter an \textit{indirect} left turn.

SUMO's default model only provides cyclists with the \textit{direct} left turn behavior, thus significantly limiting the simulation's degree of realism (see also \cref{fig:analysis_im_traj_default}).

Taking a closer look at real world intersections in the SimRa dataset revealed that there are two distributions of left-turn behavior in practice.
In the first, the \textit{indirect} path is chosen with a probability of 57 $\pm$7\%, while on intersections following the second, almost all cyclists choose the \textit{indirect} path.
Randomly investigating intersections of both types revealed that the first type has no specific characteristics while the second type actively encourages cyclists to indirect turns through the design of the intersection, e.g., by having a traffic island in the center.
Since such information cannot be identified in \ac{OSM} data reliably and in an abstract way, we will consider only the first distribution in the following.

SUMO and real world data differ decisively in all metrics considered, namely acceleration, velocity, and left-turn behavior at intersections.
However, these three metrics are crucial for realistically simulating bicycle traffic.
In the following, we try to adapt SUMO to simulate a more realistic cyclist behavior.

\section{Improving SUMO's Bicycle Simulation}
\label{sec:concept}
To improve the simulation, we propose two changes to SUMO's bicycle model:
First, the longitudinal kinematic parameters of SUMO's default bicycle model are \mbox{(re-)}parameterized based on the findings from the SimRa dataset.
Second, a novel simulation model is derived from SimRa trajectories to exclusively simulate realistic left-turn bicycle behavior at intersections based on the findings in \cref{sec:analysis}.
The latter model is referred to as the intersection model in the following.

\subsection{Longitudinal Kinematic Behavior}
\label{sec:concept_param}

In \cref{sec:analysis}, we derived maximum acceleration and maximum velocity characteristics from the SimRa dataset.
We now use them to improve the longitudinal kinematic behavior of the default SUMO bicycle model.
Contrary to the default parameterization, we use theoretical distribution functions instead of scalar values for the exposed kinematic parameters.
This enables the model to produce more realistic bicycle simulation results since the heterogeneity of real world cycling styles is reflected.

We derive the theoretical distributions by aggregating the respective features from \cref{sec:analysis_acceleration,sec:analysis_velocity}.
For this, we rely on the \textit{law of large numbers} which states that the average of the results obtained from a large number of trials of the same experiment eventually converges to its true expected value~\cite{etemadi1981elementary}.
In the context of this work, this means that individual rides do not matter but that the aggregates of multiple rides will converge towards their actual expected value given a sufficiently large number of rides.

In our analysis, we found no correlation between $a_{max}$ and $v_{max}$ and, hence, decided to use both independently in our kinematic model.
For the implementation, we used \textit{vTypeDistributions} following the results of our previous analysis and sample both distribution independently.

It should be noted that through the parameterizations with theoretical probability density functions SUMO's \textit{speedDev} parameter becomes obsolete as variance between the kinematic preferences among cyclists are already represented by the distribution function.

\subsection{Left-turn Behavior at Intersections}
\label{sec:concept_im}

To improve the degree of realism in cyclists' left-turn behavior at signaled intersections, we use an adapted version of the external intersection model (a Python script that steers cyclists via SUMO's \textit{Traffic Control Interface}) as proposed by \textcite{kaths2016integration} which is based on previously recorded real-world trajectories as their guidelines for cyclists across a single predefined intersection.
Our approach algorithmically synthesizes the cyclists' trajectories (i.e., their respective guidelines across the intersection) for any regular four-way intersection and can therefore be seen as a step towards a more universal solution.

The left-turn maneuver distribution, as we call it, specifies the probability of the cyclists choosing either the \textit{direct} or the \textit{indirect} path to cross the intersection.
For this, we use the distribution derived in \cref{sec:analysis_pathfinding} as the default for our intersection model.
Users, however, can adjust the distribution if desired or needed for their specific purposes (see also the exception cases in \cref{sec:analysis_pathfinding}).

\section{Evaluation}
\label{sec:eval}
In this section, we evaluate our new approach from \cref{sec:concept} by comparing it to SUMO's default simulation model and the real-world data taken from the SimRa data set.
We start by analyzing acceleration (\cref{sec:eval_accel}), velocity (\cref{sec:eval_velocity}), and left-turn behavior at intersections (\cref{sec:eval_turn}) before evaluating the combination of all model extensions (\cref{sec:eval_combined}).
Please note: While it may appear obvious that using the SimRa data set for both parameterization and evaluation should lead to perfect results, this is not the case as our extensions are subject to the design restrictions imposed by SUMO

\subsection{Simulation Setup}
\label{sec:eval_setup}

\begin{figure}
\includegraphics[width=\columnwidth]{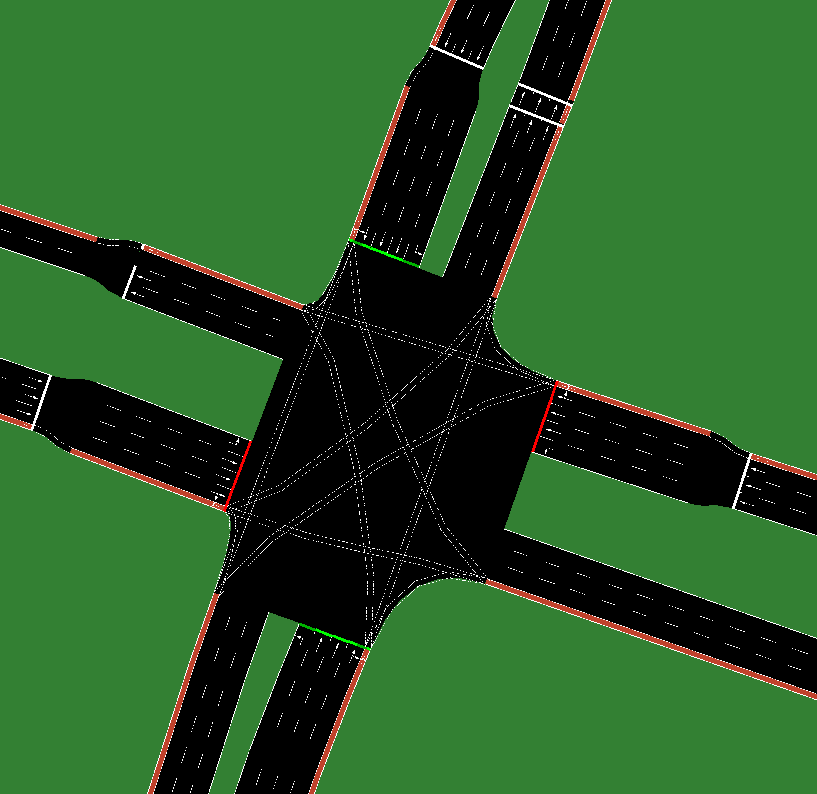}
\caption{
    Excerpt from the \textit{Mehringdamm} scenario in SUMO.
    The scenario was created using OSM data only.
}
\label{fig:mehringdamm_sumo}
\end{figure}

As SUMO users can import real-world scenarios from OSM data, simulation results can be compared to real-world data and thus be evaluated.
For our evaluation, we chose specific traffic scenarios that are representative and likely to showcase both strengths and weaknesses of our extensions.
For evaluating the longitudinal behavior (acceleration and velocity), we chose urban traffic scenarios with long straight sections.
As an example location, we use \textit{Oranienstraße} in Berlin.
For evaluating left-turn behavior, we chose compact scenarios around signaled intersections.
For this, we study three intersections in Berlin, namely at \textit{Mehringdamm} (see also \cref{fig:mehringdamm_sumo}), \textit{Warschauer Straße}, and \textit{Alexanderstraße}.

For our evaluation, we use SUMO version 1.7.0 and a step size of \SI{1}{\s} in simulations.
The SUMO default results are obtained with SUMO's \textit{vType} \texttt{Bicycle} for cyclists.
Thus, the maximum acceleration and maximum veocity are scalars and set to \SI{1.2}{\m\per\s\squared} and \SI{5.6}{\m\per\s} respectively.

\subsection{Acceleration}%
\label{sec:eval_accel}

\begin{figure}
\includegraphics[width=\columnwidth]{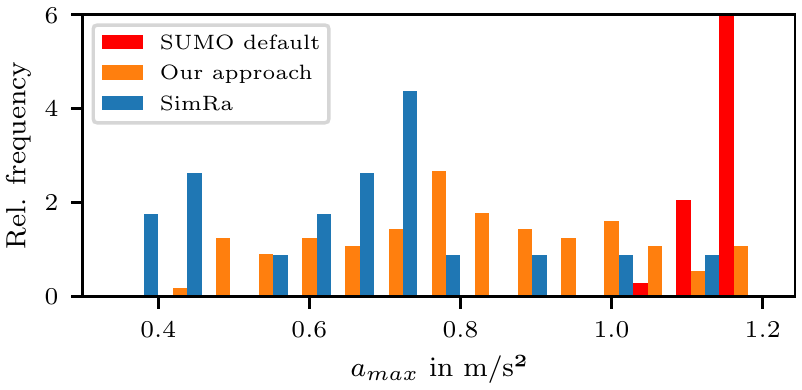}
\caption{
    Histogram of SUMO's, SimRa's, and our approach's observed maximum accelerations inside the example \textit{Oranienstraße} scenario.
    While the maximum accelerations are heterogeneously distributed in the real-world data and our approach, the default values are clustered.
    Here, the observed accelerations deviate from the configured default value (1.2m/s\textsuperscript{2}) due to traffic effects inside the simulation.
}
\label{fig:eval_acc}
\end{figure}

\cref{fig:eval_acc} shows the empirical distributions of the maximum acceleration among cyclists inside a specific simulation scenario and the corresponding real world data.
It is evident that real-world acceleration maneuvers show heterogeneous maximum rates of acceleration.
Apparently, the default parameterization is not suitable to describe this acceleration behavior among cyclists, as it provides homogeneous maximum acceleration rates within the simulation.
Our new parameterization is significantly closer to the real-world behavior in the SimRa data set with its highly heterogeneous behavior across cyclists.
That our new parameterization is not a perfect fit indicates that there are probably additional influence factors, e.g., the traffic density or the the weather situation, not covered in our kinematic model which aggregates data from all SimRa rides.

\subsection{Velocity}%
\label{sec:eval_velocity}

\begin{figure}
\includegraphics[width=\columnwidth]{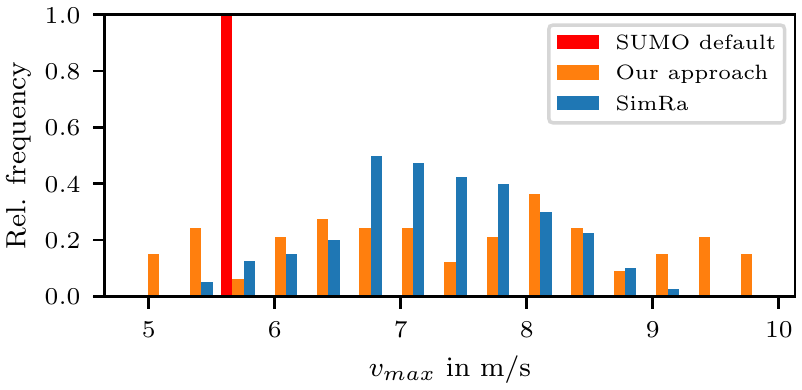}
\caption{
    Histogram of SUMO's, SimRa's, and our approach's observed maximum velocities inside the example \textit{Oranienstraße} scenario.
    While the maximum velocities are heterogeneously distributed in the real-world data and our approach, the default values are clustered.
}
\label{fig:eval_velo}
\end{figure}

\cref{fig:eval_velo} shows the empirical distributions of the cyclists' maximum velocities in the same simulation and the real-world scenario.
As with maximum acceleration rates, maximum velocities vary widely among real-world cyclists.
Once more, the default parameterization is not able to reflect this characteristic.
Our new parameterization is thus significantly closer to the real-world behavior of cyclists.
As for acceleration behavior, the fact that our new parameterization is not a perfect fit to the real-world data indicates that there are likely to be additional influence factors not captured in our model.

\subsection{Left-turn Behavior at Intersections}%
\label{sec:eval_turn}

\begin{figure}
\includegraphics[width=\columnwidth]{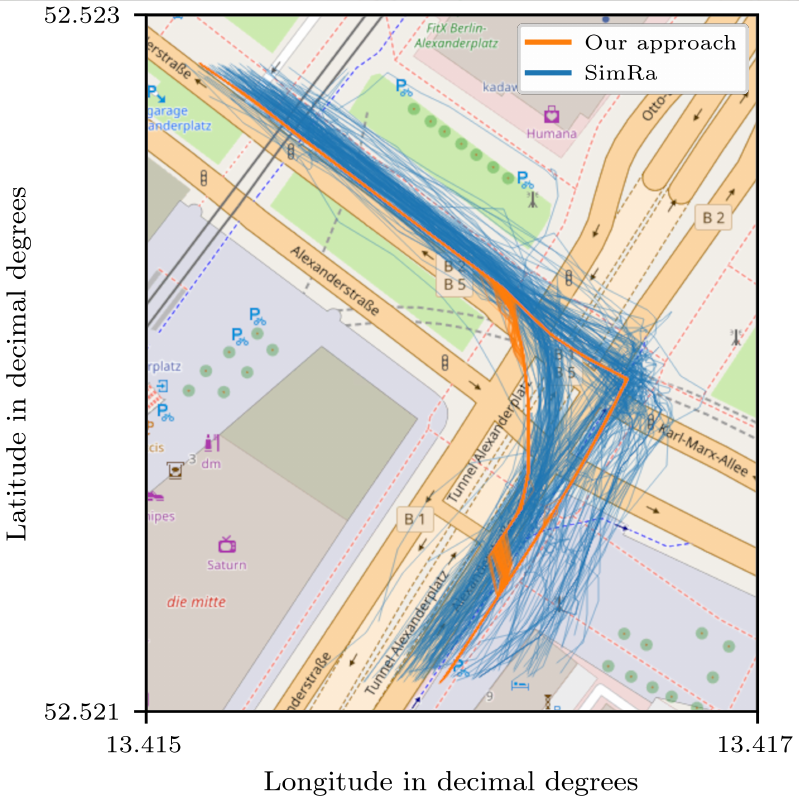}
\caption{
    Qualitative comparison between the results of our approach for the intersection model and real-world data given by SimRa for the intersection between Alexanderstraße and Karl-Marx-Allee in Berlin.
    SimRa shows two distinct left-turn paths (i.e., a direct and an indirect one), which are also modeled by our approach.
}
\label{fig:eval_im_traj_new}
\end{figure}

As shown in \cref{fig:eval_im_traj_new}, which shows the intersection between Alexanderstraße and Karl-Marx-Allee in Berlin, the 2D trajectories produced by the new intersection model converge towards the trajectories of the SimRa data set.
While the trajectories produced by SUMO's default bicycle model only offer direct turns (we omit those in the figure), the new model is significantly closer to real-world intersection behavior of cyclists.

\subsection{Combining Intersection Model and Kinematic Extensions}%
\label{sec:eval_combined}

To achieve a holistic comparison between SUMO's default bicycle model and our new approach, we measure the durations of left-turn maneuvers at multiple intersections and compare the empirical distributions of these measurements.
To specifically monitor the impact of our changes, we do not include any ride time before or after the intersection in the measurements.

Based on this, we identified the following three findings:


\begin{figure}[t]
\includegraphics[width=\columnwidth]{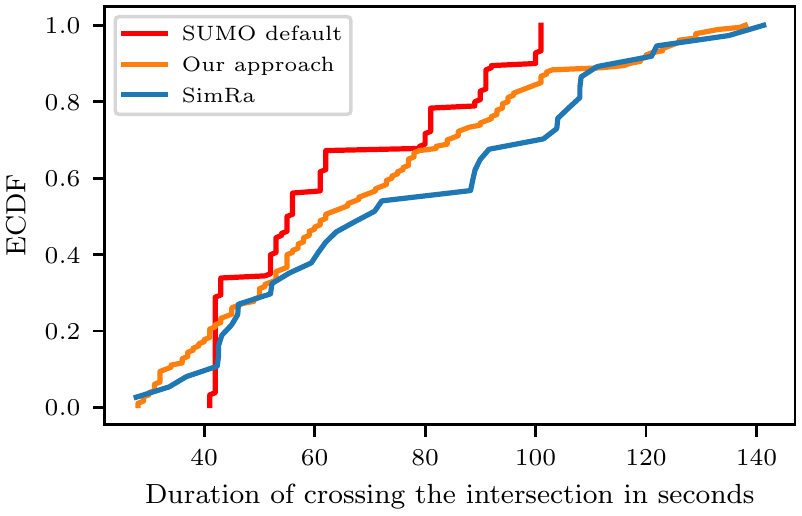}
\caption{
    ECDFs of the measured durations for crossing the scenario \textit{Warschauer Straße}.
    It is apparent that our approach outperforms SUMO's default as the measured durations converge towards the real-world data.
}
\label{fig:im_warschauer}
\end{figure}

First, our new approach outperforms the default at most intersections, as its measured durations converge with real data, see for example \cref{fig:im_warschauer}.
Especially when given the option to use the \textit{indirect} path, cyclists take longer to cross an intersection as they need to stop at an additional traffic light.
This is consistent with real-world data as we find it in the SimRa dataset at multiple intersections.


\begin{figure}[t]
\includegraphics[width=\columnwidth]{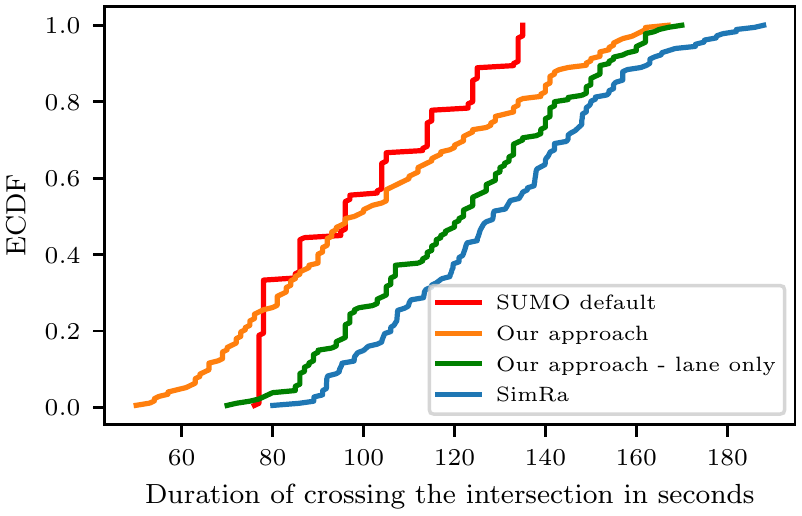}
\caption{
    ECDFs of the measured durations for crossing the scenario \textit{Mehringdamm}.
    The results when using our new approach are only slightly more realistic than when using the standard SUMO model.
    However, when the \textit{direct} path is blocked for cyclists, the simulation results outperform the default approach.
}
\label{fig:im_mehringdamm}
\end{figure}

Second, in some cases, we were able to improve our results by adjusting the left-turn behavior distribution following the second distribution discussed in \cref{sec:analysis_pathfinding}.
The ``lane only'' results in \cref{fig:im_mehringdamm} were achieved by prohibiting cyclists from using the \textit{direct} path.
Obviously, it takes much longer for cyclists to cross the intersection than SUMO's default simulation model suggests.
When examining SimRa trajectories at this particular intersection, almost all cyclists chose the \textit{indirect} path as the infrastructure guides cyclists to do so.
Hence, manually adjusting the left-turn behavior distribution for such intersections is crucial.

\begin{figure}[t]
\includegraphics[width=\columnwidth]{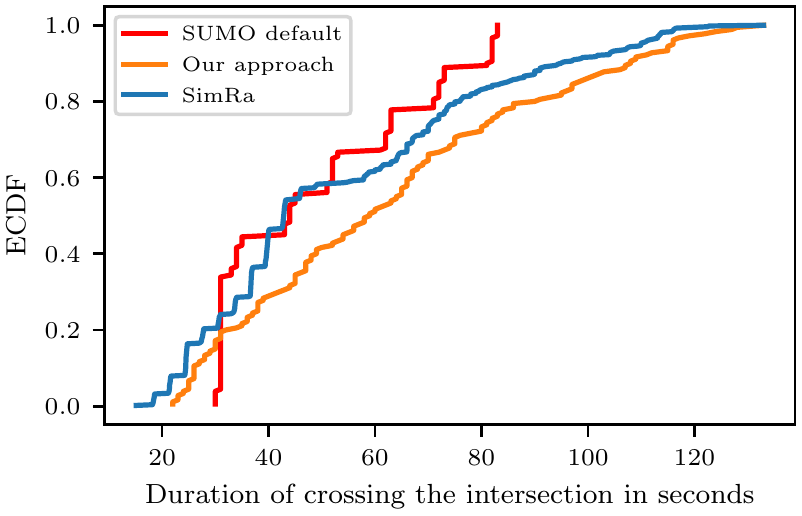}
\caption{
    ECDFs of the measured durations for crossing the scenario \textit{Alexanderstraße}.
    Here, our new approach is not more precise than SUMO default behavior.
}
\label{fig:im_alex}
\end{figure}

Third, at a few intersections, the results of our approach do not yet sufficiently reflect real-world bicycle behavior (see \cref{fig:im_alex}).
We discuss possible reasons for this in \cref{sec:disc}.

\section{Discussion}
\label{sec:disc}
Overall, the results presented in this paper show a significant improvement over the state-of-the-art.
Nevertheless, they still have a number of shortcomings.
In this section, we discuss the inherent limitations of our approach in general (\cref{sec:method}) as well as problems resulting from the SimRa dataset as our ground truth data (\cref{sec:problem_simra}).

\begin{figure}
\includegraphics[width=\columnwidth]{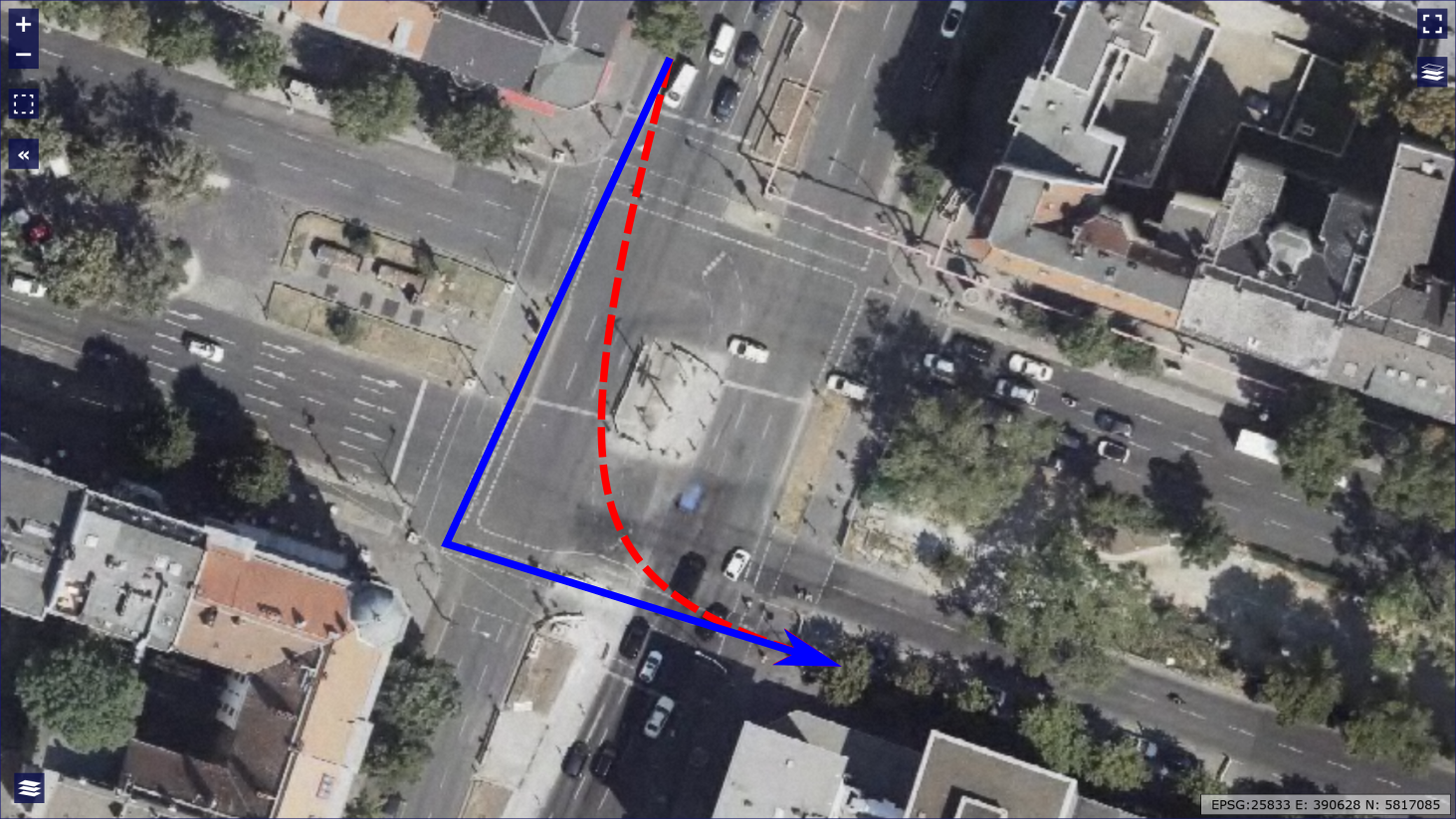}
\caption{Intersection Mehringdamm/Gneisenaustraße: a traffic island obstructs the direct turn path (dashed line) and, thus, makes the indirect path (solid line) more likely to be used.}
\label{fig:mehringdamm_traffic_island}
\end{figure}

\subsection{Methodological Challenges}
\label{sec:method}

Our initial assumption was that the behavior of cyclists in a single intersection cannot be generalized to all intersections~\cite{kaths2016integration} but that the average behavior across a large number of intersections will be close enough to cyclists' behavior at arbitrary intersections.
This seems to be true only for a (relatively large) subset of intersections -- apparently, the intersection behavior of cyclists is more heterogeneous than expected.
We believe that this is due to the fact that we averaged across all intersections in our dataset whereas there are apparently different classes of intersections that we did not account for.

Primarily, the intersection design is likely to have a strong impact:
Consider the example in \cref{fig:mehringdamm_traffic_island} where a traffic island partially blocks direct left turns and where markings on the ground suggest indirect turns.
As another example, the intersection Bismarckstraße/Leibnizstraße had no direct left turns in the SimRa dataset.
In this intersection, the reason would be that cyclists legally have to use a bike lane.
When using that bike lane, a direct left turn would require cyclists to first pass through a row of parked cars, then to cross four car lanes of a major street before being able to turn left.

Aside from that, other possible influence factors include the amount and velocity of traffic (higher numbers of cars or faster cars can be expected to lead to more indirect turns), gender and age group distributions of cyclists in the respective intersections, as well as weather and light conditions or the grade of the street.
In future work, we plan to explore these possible influence factors, focusing on the intersection design which we deem to have the strongest impact.

Another problem results from inaccuracy in the dataset used:
GPS and motion sensors of smartphones provide only imprecise insights into actual ``micro''-behavior of cyclists.
Using a broad group of cyclists as input will always result in this limitation which we tried to overcome based on preprocessing and filtering of the SimRa dataset.
Alternatives would be additional sensors (especially cameras) on bicycles or on intersections as in~\cite{kaths2016integration}.
These, however, have the inherent limitation that they will either limit the number of bicycles producing data or the number of intersections covered.

\subsection{Dataset Choice as Ground Truth}
\label{sec:problem_simra}

In this paper, we used the SimRa datasets as input for our analysis as it is, to the best of our knowledge, the first public dataset comprising a large number of rides that actually publishes individual rides in an anonymized but non-aggregated form.
We need to keep in mind, however, that SimRa was designed for a different purpose:
For example, the SimRa app records motion sensors at 50Hz but only persists every fifth value of a moving average over 30 values.
While this suffices for detecting near miss incidents~\cite{karakaya2020simra}, it further limits the resolution of motion data (and thus any conclusions we can draw from that).
Furthermore, the SimRa data which we used were recorded over a period of 1.5 years.
During such as long period of time, physical changes to the bicycle infrastructure (both temporary and permanent) will occur, thus, adding additional noise to the data.

Finally, SimRa relies on crowdsourcing as a data collection method which often leads to participation inequality.
As a result, individual users will be overrepresented in some intersections and street segments and not represented in others.
Furthermore, based on the data collection method using smartphones, the user group of SimRa is likely to have a slight gender bias towards males and an age group bias towards cyclists between the ages 20 and 50.
These biases will, of course, be reflected in our analysis results and cannot be compensated unless other cycling datasets become available in non-aggregated form.

\subsection{Generalizability}
\label{sec:problem_general}

Although the SimRa dataset contains rides from almost 100 regions, we only considered rides from Berlin for our data analysis and the development of cyclists' left turn behavior.
The main reason for this is that almost half the rides are from the Berlin region~\cite{karakaya2022cyclesense}.
To derive both turn model and acceleration behavior, however, we need data from "dense" street segments with many rides which are hence only available for Berlin at the moment.
This is further aggravated by the fact that the preprocessing step (see \cref{sec:analysis}) further reduces the number of eligible rides and with that the number of intersections with sufficient left-turn maneuvers.
Nevertheless, this does not pose a threat to the generalizability of our findings -- at least for Germany -- as traffic infrastructure guidelines throughout Germany are standardized.
Furthermore, adjacent countries such as Austria often also have comparable infrastructure.
We hence believe that our findings also apply to at least these countries.

\section{Conclusion}
\label{sec:conclusion}
Increasing the modal share of cyclists to provide health benefits, alleviate traffic congestion, and reduce air pollution requires significant planning efforts of city planners and traffic engineers towards an improved cycling infrastructure.
For this, city planners often rely on the open source simulation platform SUMO to study the effects of infrastructure changes before implementing them on the streets.
Likewise, research on V2X-based safety systems for cyclists often relies on SUMO for evaluation.
Unfortunately, SUMO cyclists are either modeled as slow cars or as fast pedestrians, neither of which is overly realistic.

In this paper, we used the recently published SimRa dataset, which to our knowledge is the first public dataset providing detailed insights into a large number of individual cyclists' rides, to improve SUMO's cyclist model.
For this, we derived acceleration and velocity behavior and reparameterized the SUMO cyclist model.
As a SUMO extension, we also developed a new intersection model describing left-turn behavior of cyclists in four-way intersections.
While our work significantly improved the existing cyclist model, it is not as realistic as we wanted it to be.
We, hence, discussed a number of research directions which we plan to explore in the near future.

\printbibliography
\end{document}